\documentclass[a4paper,12pt]{article}
\pdfoutput=1
\usepackage{amsmath}
\usepackage{amssymb}
\usepackage{amsfonts,amsthm,dcolumn}
\usepackage{caption}
\usepackage{graphicx}
\usepackage{subfigure}
\usepackage{color}
\usepackage{url}

\RequirePackage[numbers,sort&compress]{natbib}
\RequirePackage{color}
\RequirePackage[colorlinks=true
,urlcolor=blue
,anchorcolor=blue
,citecolor=blue
,filecolor=blue
,linkcolor=blue
,menucolor=blue
,linktocpage=true
,pdfproducer=medialab
,pdfa=true
]{hyperref}

\DeclareGraphicsExtensions{.pdf,.png,.jpg,.jpeg,.eps}

\def\nn{\nonumber}
\def\mco{\mathcal{O}}

\begin{document}
\begin{titlepage}
\thispagestyle{empty}
\begin{flushright}
ICTS/2013/16
\end{flushright}
\bigskip
\begin{center}
\noindent{\Large \textbf{Dissipative Nonlinear Dynamics in Holography}}\\
\vspace{2cm} \noindent{
{\Large Pallab Basu\footnote{email:\href{mailto:pallab.basu@icts.res.in}{pallab.basu@icts.res.in}},
Archisman Ghosh\footnote{email:\href{mailto:archisman.ghosh@icts.res.in}{archisman.ghosh@icts.res.in}}}}\\
\vspace{1cm}
{\it 
Department of Physics and Astronomy, \\ University of Kentucky, Lexington, KY 40506, USA \\ \vspace{.5cm} 
International Centre for Theoretical Sciences, \\ Tata Institute of Fundamental Research, \\ Bangalore 560012, INDIA \\}
\end{center}
\vspace{0.3cm}
\begin{abstract}
We look at the response of a nonlinearly coupled scalar field in an asymptotically $AdS$ black brane geometry and find a behavior very similar to that of known dissipative nonlinear systems like the chaotic pendulum. Transition to chaos proceeds through a series of period-doubling bifurcations. The presence of dissipation, crucial to this behavior, arises naturally in a black hole background from the ingoing conditions imposed at the horizon. $AdS$/CFT translates our solution to a chaotic response of $\mco$, the operator dual to the scalar field. Our setup can also be used to study quenchlike behavior in strongly coupled nonlinear systems.
\end{abstract}
\end{titlepage}
\newpage
\section{Introduction}

Over the last few years $AdS$/CFT \cite{Maldacena:1997re,Gubser:1998bc,Witten:1998qj,Aharony:1999ti} has been successfully applied to various problems inspired by condensed matter systems \cite{Hartnoll:2011fn}. One kind of system which is of much interest, is one which is driven or quenched externally by force \cite{Polkovnikov:2010yn}. In the field theoretic context it translates to creating a disturbance by turning on a time-dependent source for some operator. In general, in a strongly coupled field theory, it is difficult to study the driven problems directly by employing analytic or even numerical methods. Holography gives an alternative avenue to this strongly coupled physics. Holography maps strongly coupled field theoretic problems to gravitational problems involving Einstein's equations and additional equations governing the dynamics of dual bulk fields. The resulting differential equations, classical in nature, can be studied numerically or analytically. Various authors have discussed the issue of 
driven systems in the context of holography \cite{Bhattacharyya:2008ji,Basu:2011ft,Buchel:2012gw,Basu:2012gg,Bhaseen:2012gg,Buchel:2013lla,Mandal:2013id, Gao:2012aw}. For example, the problem of slow quench has been discussed in the simplest of setups in \cite{Basu:2012gg}. In this paper, we discuss the problem of nonlinear dynamics in the same spirit in the context of holography.

We start with a simple model of a scalar field in an $AdS$ black hole geometry \cite{Iqbal:2010eh}. We drive the system by introducing a time-dependent source for the scalar operator on the boundary. Without any nonlinearity, the problem boils down to the well-known calculation of the scalar two-point function in an $AdS$ black hole \cite{Son:2007vk}. Here we would like to understand how nonlinearity changes this simplistic behavior. There can be two different ways to approach this problem; one is to consider the gravitational backreaction \cite{Buchel:2012gw} of the scalar field and the other is to turn on the scalar self-interaction. Here we choose the second approach and consider the scalar field as a probe field with negligible gravitational backreaction. One can think of the scalar as coming from some brane field. The question we ask here is what kind of steady-state behavior, if any at all, do we observe. Our system can be thought of as a driven diffusive system like a forced damped pendulum. In our case, 
absorption at the black hole horizon introduces a diffusive element. Not unexpectedly, at small values of the nonlinearity parameter the effect of nonlinearity is negligible and we get a steady-state solution. However, as we increase nonlinearity gradually we see period doubling and novel solutions in the bulk with a response at frequencies lower than that of the driving frequency. As we increase nonlinearity more signatures of nonlinear dynamics are seen as chaotic motion sets in\footnote{For other examples of chaotic dynamics in the context of string theory see  \cite{Zayas:2010fs,Basu:2011dg,Basu:2011di,Basu:2012ae,Barbon:2011pn}.}. 

Chaos, a behavior seen in many nonlinear systems, lacks a precise definition but can be roughly defined as an exponential sensitivity of the dynamics to its initial conditions \cite{Hilborn}. Chaos is more frequently characterized by the qualitative behavior of the system in the chaotic regime (which on the other hand is quite independent of the initial conditions), or by the route the system takes to chaos as a control nonlinearity parameter is varied. In various complex systems, many different routes to chaos can be seen -- period-doubling bifurcations, quasiperiodicity, intermittency, etc. Quasiperiodicity and intermittency are characteristic of systems with multiple natural frequencies and systems with higher degrees of freedom respectively. We restrict our attention to the period-doubling route of transition to chaos.

The rest of this paper is organized as follows. In Section~\ref{sec:chaos}, we review some of the features seen in simple nonlinear dissipative systems \cite{Hilborn,Ott} by taking an example of the forced damped pendulum. In Section~\ref{sec:setup}, we set up our problem and obtain the equations of motion. We solve the equations of motion numerically and in Section~\ref{sec:num-sol} we display the results for the static case and with time-dependent boundary conditions relevant to both chaotic behavior and quenches.

\section{Chaos in the driven damped pendulum}
\label{sec:chaos}
A classic example\footnote{The reader familiar with period doubling bifurcations and related aspects of dynamical systems may proceed directly to Section~\ref{sec:setup}.} of the nonlinear dissipative systems is a forced damped pendulum described by the equation of motion,
\begin{align}
\label{eq:pen-eom}
\ddot{\theta}+\gamma\dot{\theta}+\Omega^2\sin{\theta}=F(t)\,.
\end{align}
Here $\Omega$ is the natural frequency of the pendulum and $\gamma$ is the damping coefficient. The nonlinearity arises due to the presence of the ``$\sin{\theta}$'' term. The driving is typically given by sinusoidal forcing with amplitude $A$ and frequency~$\Omega$:
\begin{align}
\label{eq:pen-forcing}
F(t)=A\sin\,\Omega{}t\,.
\end{align}

\begin{figure}[h!]
\centering
\subfigure[Phase portrait showing $4$-cycle.]{
\includegraphics[scale=0.8]{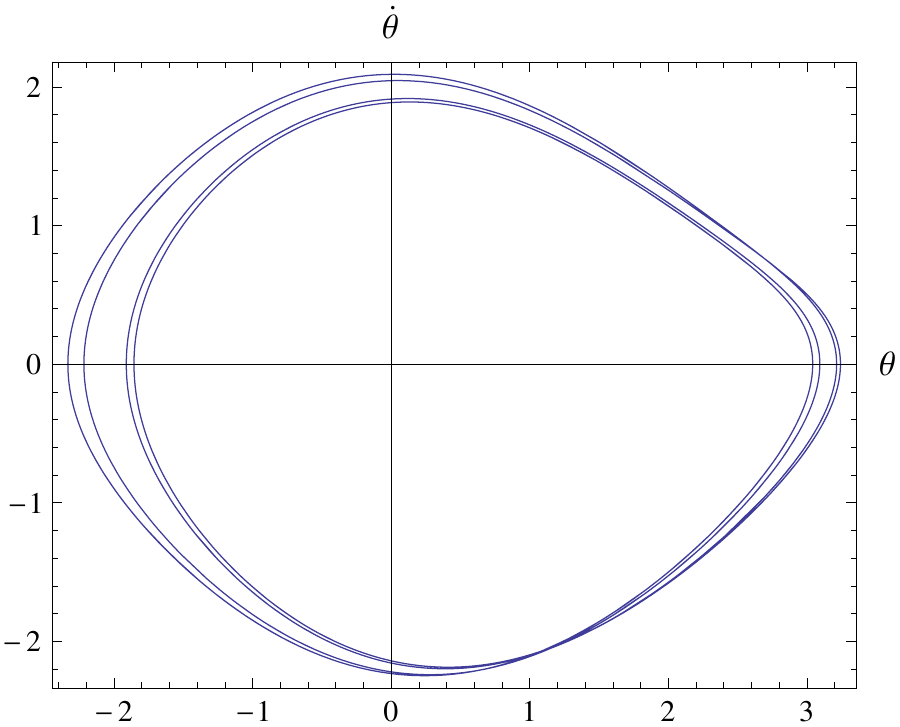}
\put(-48,140){\tiny{$\Omega=2/3$,}}
\put(-50,130){\tiny{$A=1.081$}}
\label{fig:pendulum_4-cycle}
}
\subfigure[Bifurcation diagram.]{
\includegraphics[trim=0cm 0.5cm 0cm 0cm, clip=true, width=0.45\textwidth, height=0.32\textwidth]{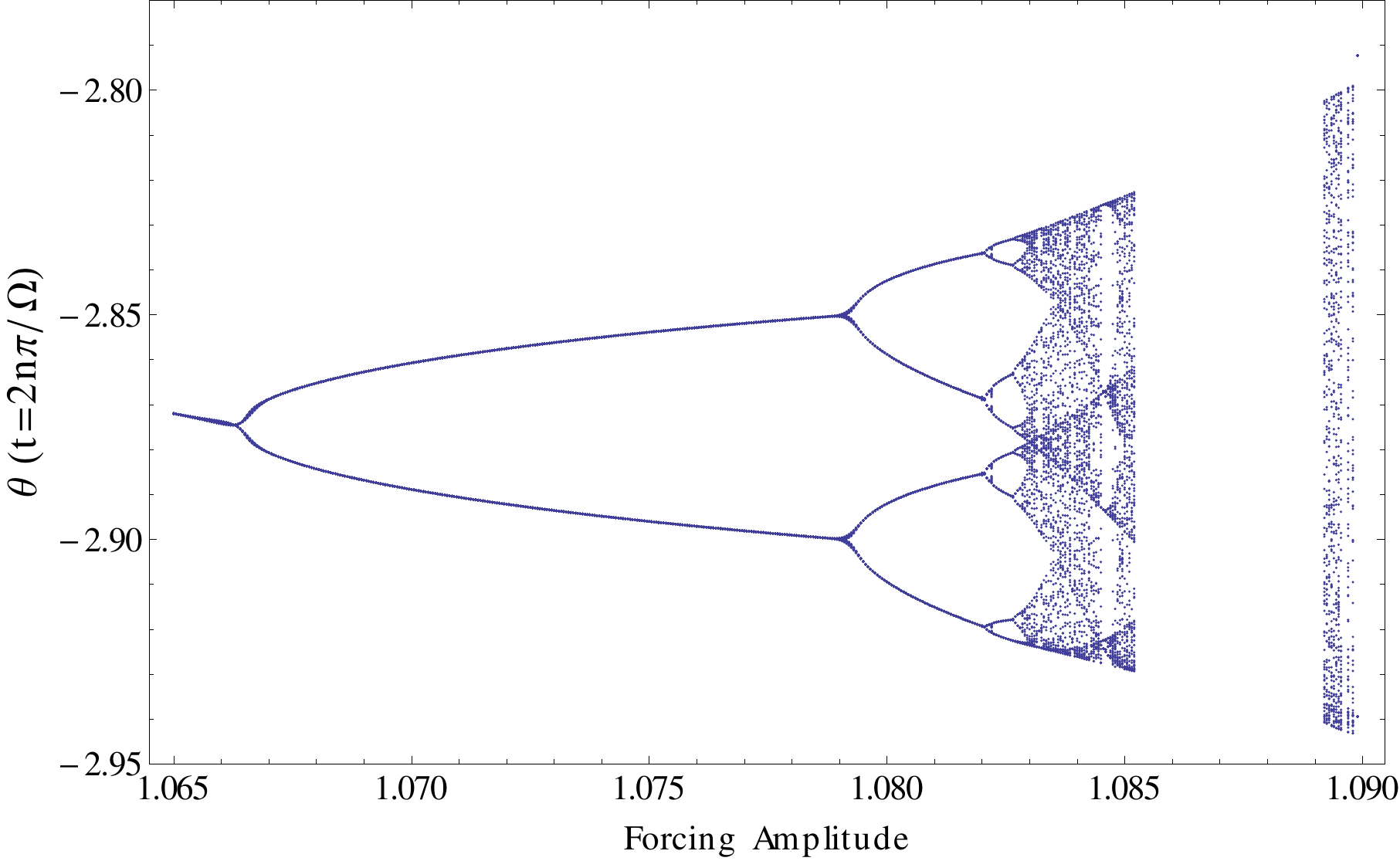}
\put(-183,140){\tiny{$\Omega=2/3$}}
\label{fig:pendulum_bifdiag}
}
\caption{Nonlinear response of the pendulum (with $\Omega=1$, $\gamma=0.5$). Fig.\,\ref{fig:pendulum_4-cycle} shows the phase portrait for $\Omega=2/3$, $A=1.081$. We see four lobes corresponding to a 4-cycle. Fig.\,\ref{fig:pendulum_bifdiag} shows the bifurcation diagram, typical to this route of transition to chaos.}
\label{fig:pendulum_results}
\end{figure}

For early times, there is a usually {\em transient behavior}. We focus on the behavior at late times. This behavior is qualitatively different for different driving parameters ($A,\Omega$). For small amplitudes $A$, the late-time behavior is insensitive to the precise initial conditions we begin with. All trajectories in the phase space converge to a periodic {\em attractor} or a {\em limit cycle}. For very small amplitudes, the response frequency is at the forcing frequency $\Omega$ and is roughly sinusoidal. The {phase portrait} is a closed curve, roughly an ellipse; we call this limit cycle a {\em 1-cycle}. As the forcing amplitude is increased, the phase space ellipse gets distorted due to nonlinearity. However at a particular value of the amplitude the ellipse abruptly splits into two -- the alternate crests and troughs begin to have slightly different heights. The response is thus no longer at frequency $\Omega$ now, but $\Omega/2$. This 
transition is known as {\em period-doubling}. As mentioned before, the phase portrait for the motion is now not one but two intersecting loops -- this limit cycle is a {\em 2-cycle}. As the forcing amplitude is increased further, we get subsequent period-doubling more and more rapidly to {\em 4-cycles} [Fig.\,\ref{fig:pendulum_4-cycle}], {\em 8-cycles} and so on. We eventually reach a regime where there is no periodicity at all and there is no closed limit cycle. In this regime the response is chaotic -- the aperiodicity is accompanied by the fact that the system is exponentially sensitive to its initial condition. 

An efficient way to represent the period-doubling transitions is to plot what is known as the {\em bifurcation diagram} [Fig.\,\ref{fig:pendulum_bifdiag}]. For each value of the control parameter (amplitude here), one looks at all the values of $\theta$ or $\dot{\theta}$ attained after intervals of multiples of the forcing period -- so this is a stroboscopic sampling of the phase space. For a 1-cycle only 
one value is attained, for a 2-cycle two different values, for a 4-cycle four values etc., and the period-doubling transition from one to the other is clearly seen as a splitting or a {\em bifurcation} in this plot.

The period-doubling bifurcation route of transition to chaos is generic to nonlinear systems with a low number of degrees of freedom including even very simple examples like the tent or logistic map \cite{Hilborn,Ott}. The existence of period-doubling bifurcations can for all practical purposes be treated as a positive signature of chaotic behavior in the system. Some other positive indicators of chaos are the presence of $3$ and other odd cycles past the chaotic regime, which we will also see in our system below.

\section{Setup}
\label{sec:setup}
\noindent

We consider a black brane geometry in asymptotically $AdS_{d+1}$ described by the metric in
ingoing Eddington-Finkelstein-like coordinates in which the metric becomes
\begin{align}
\label{eq:metric-v}
ds^2=-f(r)\,dv^2+2\,dv\,dr+r^2\,d{\bf x}_{(d-1)}^2\,.
\end{align}
Here the warp factor is
\begin{align}
\label{eq:f(r)}
f(r)=r^2\left(1-\frac{r_h^d}{r^d}\right)\,.
\end{align}
We add a minimally coupled scalar field $\Phi$ with a nonlinear potential term, given by the Lagrangian:
\begin{align}
\label{eq:action}
S=-\int{d^{d+1}x}\sqrt{-g} \left\{ \frac{1}{2}g^{ab}\nabla_a\Phi\nabla_b\Phi + \frac{1}{2}m^2\Phi^2 + \frac{1}{4}\lambda\Phi^4\, \right\}.
\end{align}
We work in the probe limit\footnote{It should be noted that in the non-probe case there is always a horizon formation \cite{Faulkner:2010fh,Garfinkle:2011tc}.} where the field $\Phi$ does not backreact on the geometry \cite{Iqbal:2010eh}. The $\Phi^4$ interaction leads to a nonlinear term in the equation of motion. It is useful to make a coordinate transformation  
\begin{align}
\label{eq:rho}
\rho=\int_r^\infty\frac{dr'}{f(r')}
\,,{\rm \ such\  that\ }-f(r)\partial_r\equiv\partial_\rho\,,
\end{align}
and a field redefinition $\Phi=r^{-\frac{d-1}{2}}\phi$ \cite{Arean:2010zw}. Then the full equation of motion takes the form
\begin{align}
\label{eq:eom}
2\partial_v\partial_\rho\phi-\partial_\rho^2\phi+\left[V(r)+m^2f(r)\right]\phi+\lambda{}r^{-(d-1)}f(r)\phi^3=0\,.
\end{align}
Close to the boundary, where $f(r)\approx{}r^2\approx{}\rho^{-2}$, the two solutions scale as
\begin{align}
\label{eq:scaling}
\phi(\rho\approx0)\sim\rho^{\Delta_\pm}\,, {\rm \ with\ } \Delta_\pm=\frac{1\pm\sqrt{d^2+4m^2}}{2}\,.
\end{align}
We choose to work with $m^2={(1-d^2)}/{4}$ which  is the conformal mass (it is still above the Breitenlohner-Freedman \cite{Breitenlohner:1982bm, Breitenlohner:1982jf} bound $-d^2/4$). For this value of $m^2$, $\Delta_\pm=1,0$ and
\begin{align}
\label{eq:scaling-spl}
\phi(\rho\approx0)\sim\phi_0-\rho\phi_1+\ldots
\end{align}
Near the boundary, the bulk field $\Phi$ scales as
\begin{align}
\label{eq:field-expan}
\Phi(\rho,x^\mu)=\rho^{\frac{d-1}{2}}\left\{{\rho^{\Delta_-}\phi_0(x^\mu)-\rho^{\Delta_+}\phi_1(x^\mu)}\right\}.
\end{align}
Using the standard $AdS$/CFT prescription, one interprets $\phi_1$ as the expectation value $\langle\hat{\mco}\rangle$ of the operator dual to the field $\Phi$ sourced by its boundary value $J\equiv\phi_0$. Since in our case $\Delta_+-\Delta_-=1$, $\langle\hat{\mco}\rangle$ is given simply by
\begin{align}
\label{eq:op}
\langle\hat{\mco}\rangle\equiv\phi_1=\left( -\frac{\partial\phi}{\partial{\rho}}  +\frac{\partial\phi}{\partial v} \right)_{\rho=0} \,.
\end{align}

We need to augment our equation of motion (\ref{eq:eom}) with appropriate boundary conditions. The ingoing condition at the horizon is ensured by regularity of $\phi$ at $\rho=\infty$ in ingoing coordinates. For the condition at the $AdS$ boundary at $\rho=0$, we choose a time-dependent ``driving''
\begin{equation}
\label{eq:bc}
\phi_{\rho=0}(v)=\phi_0(v)=\left\{ \begin{tabular}{l}
                               $A \sin \Omega v$ \\
                               $A \tanh \beta v$
                              \end{tabular}
\right.
\end{equation}

\begin{figure}
\begin{center}
\includegraphics[scale=0.25]{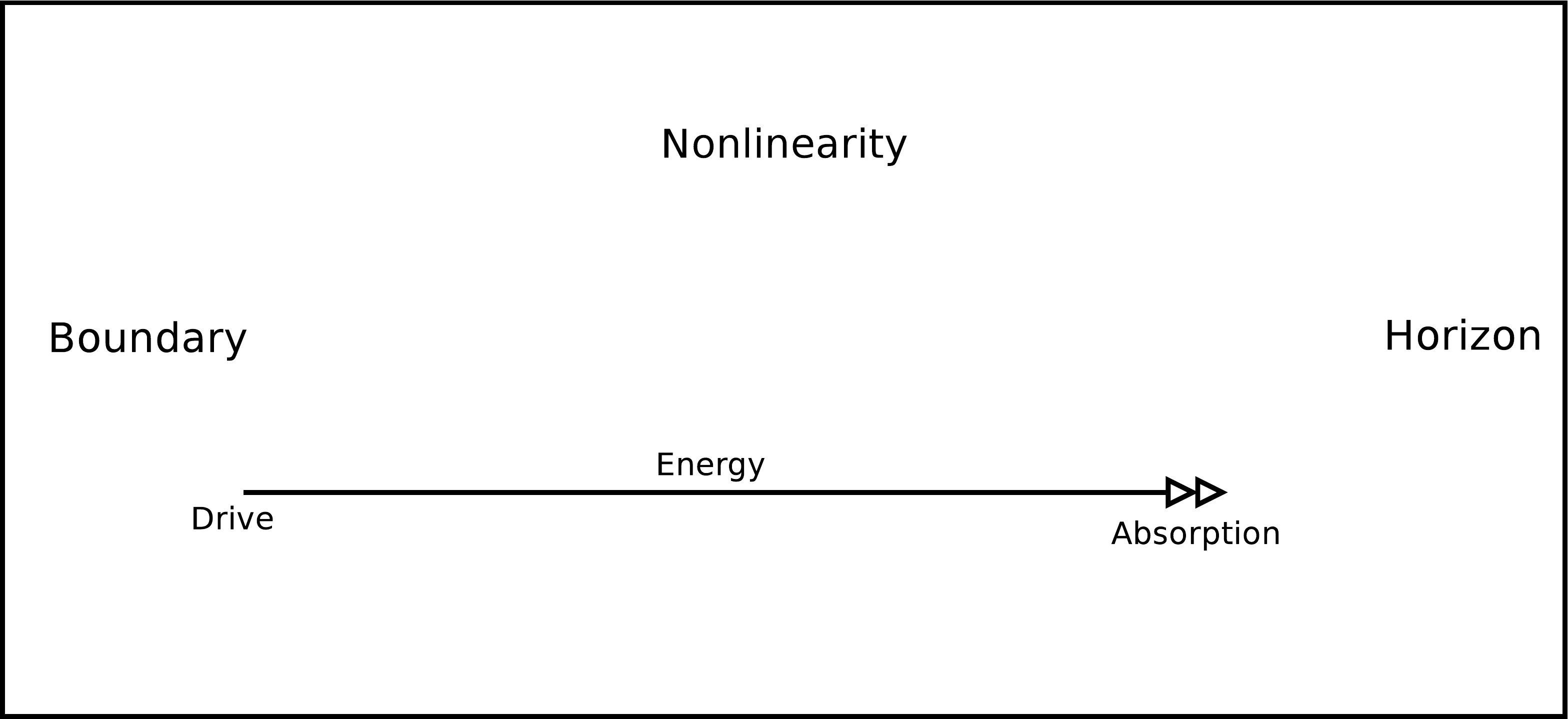}
\end{center}
\caption{Schematic diagram showing energy flow from boundary to horizon.}
\label{fig:schematic}
\end{figure}

The existence of a black hole horizon adds the element of absorption in our setup. On one side energy is pumped into the system by a time-dependent boundary condition and on the other side energy is absorbed at the horizon. In the middle, in the bulk region we have nonlinearity [Figure~\ref{fig:schematic}]. 
As we will see, using the different forms of driving, our system can be used to study forced damped dynamics as well as relaxation in quenches. 

\section{Numerical solution and results}
\label{sec:num-sol}
In the numerics we focus on $d=4$ and solve the equation of motion, Eq.(\ref{eq:eom}), numerically. Our equation is similar to a wave equation in a semi-infinite plane and hence we can use the numerical technique of method of lines\footnote{It should be noted that it is considerably difficult to solve the equations if gravity is dynamical. In that case, the position of the horizon itself becomes time dependent and one needs to apply advanced techniques like pseudospectral methods. See, for example, \cite{Pfeiffer:2002wt}.}. We discretize Eq.(\ref{eq:eom}) on a uniform grid in $\rho$ by replacing the partial derivatives by finite-difference derivatives. The regularity condition at $\rho=\infty$ is identical to a Neumann condition $\left.\partial_\rho\phi\right|_{\rho\to\infty}=0$ and this implies that $\phi$ attains a constant value for large $\rho$. We can thus cut off the domain of our problem to a large but finite value of $\rho$. It serves our purpose to choose $\rho_{\rm max}=5.5$ and $N=2500$ points on 
the grid\footnote{It is clear from Figs.\ref{fig:typical}-\ref{fig:profile} that $\phi(\rho)$ does indeed go to a constant for the $\rho_{\rm max}$ we choose.}. We solve the resulting ordinary differential equations (ODEs), which are $N$ coupled first-order differential equations in $v$, using the SUNDIALS IDA solver \cite{sundials}.

For the purpose of numerics, we choose $\lambda=1$. However any nonzero $\lambda$ can be scaled to $\lambda=1$ using $\lambda^\frac{1}{2}\phi\to\phi$. The forcing amplitude $A$ thus also serves as a nonlinearity parameter. Before going over to the dynamics of the system, we will briefly look at the time-independent solution.

\subsection{Static case}
\label{sec:static}
\begin{figure}[h!]
\centering
\subfigure[Typical profile of bulk field.]{
\includegraphics[scale=0.38]{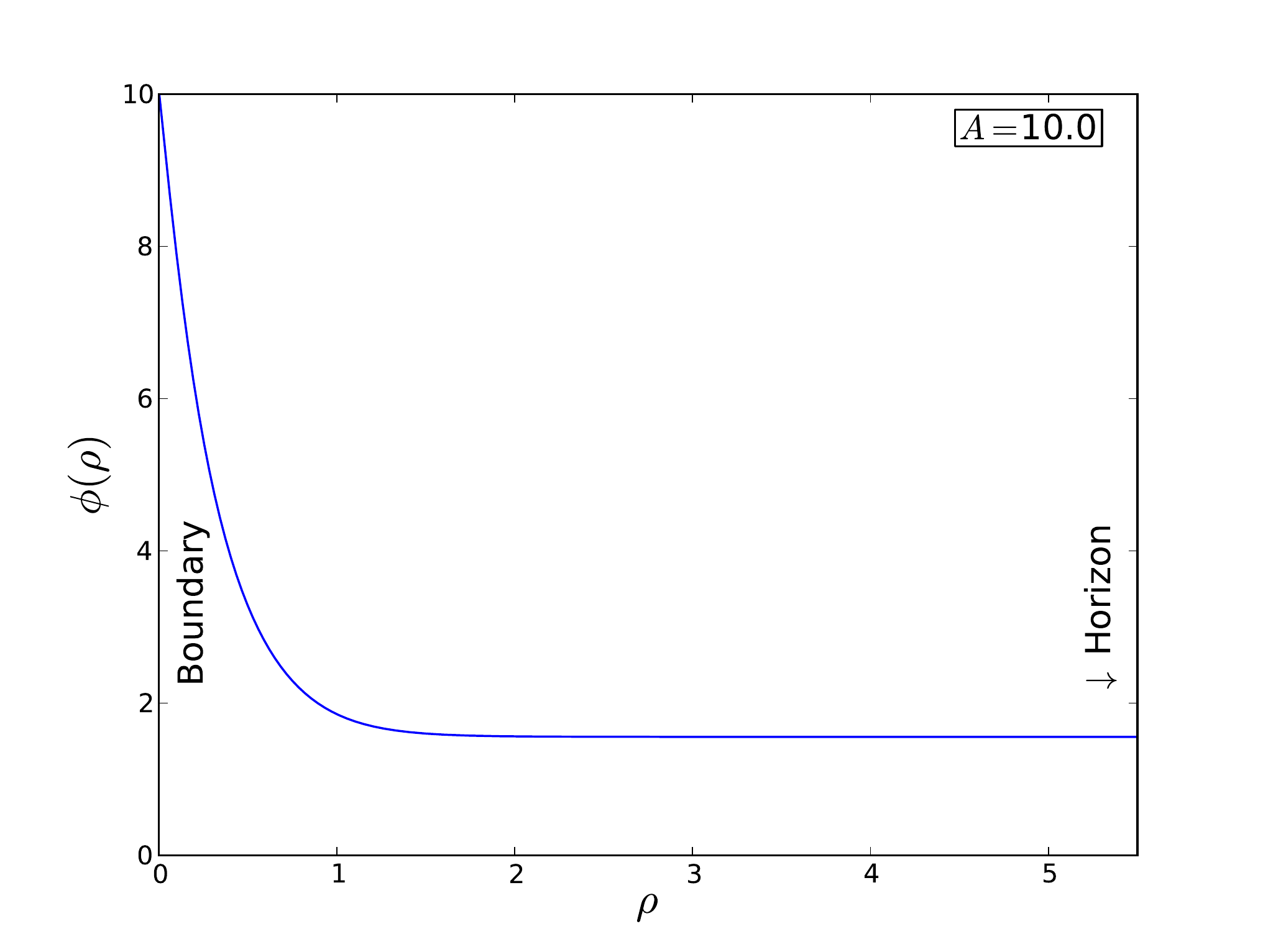}
\label{fig:typical}
}
\subfigure[Static response of boundary  operator.]{
\includegraphics[scale=0.38]{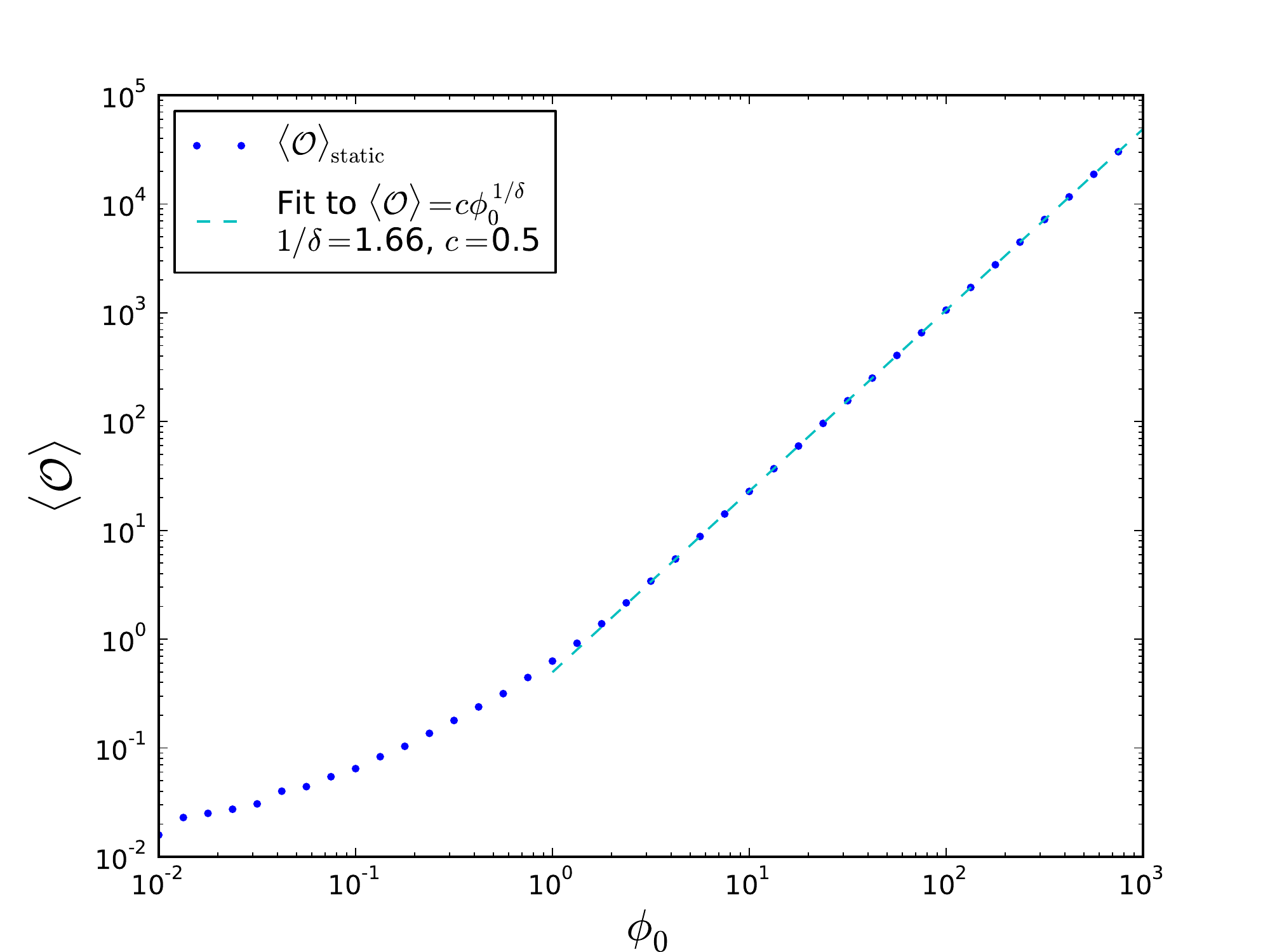}
\label{fig:scaling}
}
\caption{Static solution. Fig.\,\ref{fig:typical} shows a typical profile of the time-independent bulk field. Fig.\,\ref{fig:scaling} is a logarithmic plot of $\langle\hat{\mco}\rangle$ vs. boundary value $\phi_0$. $\langle\hat{\mathcal{O}}\rangle\sim\phi_0^{5/3}$ for large $\phi_0$.}
\label{fig:static}
\end{figure}

For the time-independent case, the equations 
are ODEs with mixed boundary conditions and the solution can be simply obtained, for example, by shooting. The results are shown in Figure~\ref{fig:static}. A typical profile of the bulk field is plotted in Fig.\,\ref{fig:typical}. Fig.\,\ref{fig:scaling} shows the response of the field theory operator expectation $\langle\hat{\mathcal{O}}\rangle$ as a function of the boundary value $\phi_0$. We see that for large values of the field (high nonlinearity), the operator scales as $\langle\hat{\mathcal{O}}\rangle\sim\phi_0^{5/3}$.

This exponent can also be obtained analytically by scaling arguments near the boundary. For small $\rho$, the potential and mass terms cancel each other for $m^2=(d^2-1)/4$ and the static equation is
\begin{align}
\label{eq:small-rho}
-\partial_\rho^2\phi(\rho)+\lambda\rho^{d-3}\phi^3(\rho)=0\,.
\end{align}
If $\phi_s(\rho)$ is a solution to this equation, then $\phi(\rho)\equiv\alpha^{(d-1)/2}\phi_s(\alpha\rho)$ is also a solution.
\begin{align}
\label{eq:Oscaling}
J\equiv\phi_{\rho=0}&=\alpha^{(d-1)/2}\phi_s(0)\,, \nn \\
\langle\hat{\mco}\rangle\equiv\phi'_{\rho=0}&=\alpha^{(d+1)/2}\phi'_s(0)\,.
\end{align}
Thus $\langle\hat{\mco}\rangle\sim{}\phi_0^{(d+1)/(d-1)}$. The near boundary scaling works because for large boundary values $\phi_0$ the field $\phi$ drops off faster with $\rho$ and its behavior is dominated by the small $\rho$ solution.

\subsection{Sinusoidal driving}

We first choose a driving $\phi_0(v)=A\sin\Omega{}v$. In the limit $A \ll 1$, the effect of nonlinearity is negligible and we have the relation 
\begin{align}
\langle \hat{\mco} \rangle (\Omega)=G(\Omega) A\,,
\end{align}
where $G(\Omega)$ is the Green's function in the black hole background.

\begin{figure}[h!]
\centering
\subfigure[Late-time bulk field (for $v=2n\pi/\Omega$).]{
\includegraphics[scale=0.38]{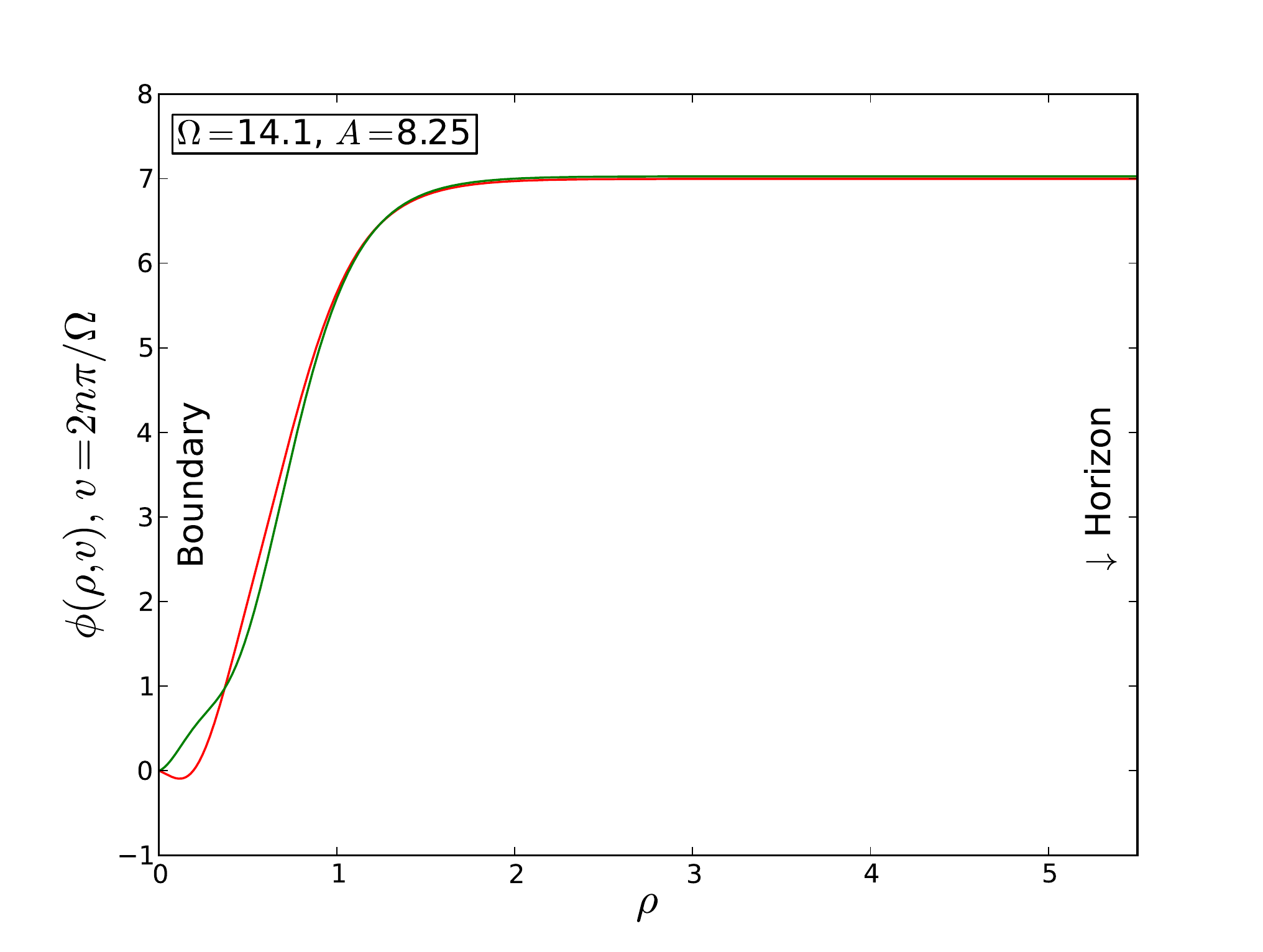}
\label{fig:profile}
}
\subfigure[Late-time behavior of $\langle\mathcal{O}\rangle(v)$.]{
\includegraphics[scale=0.38]{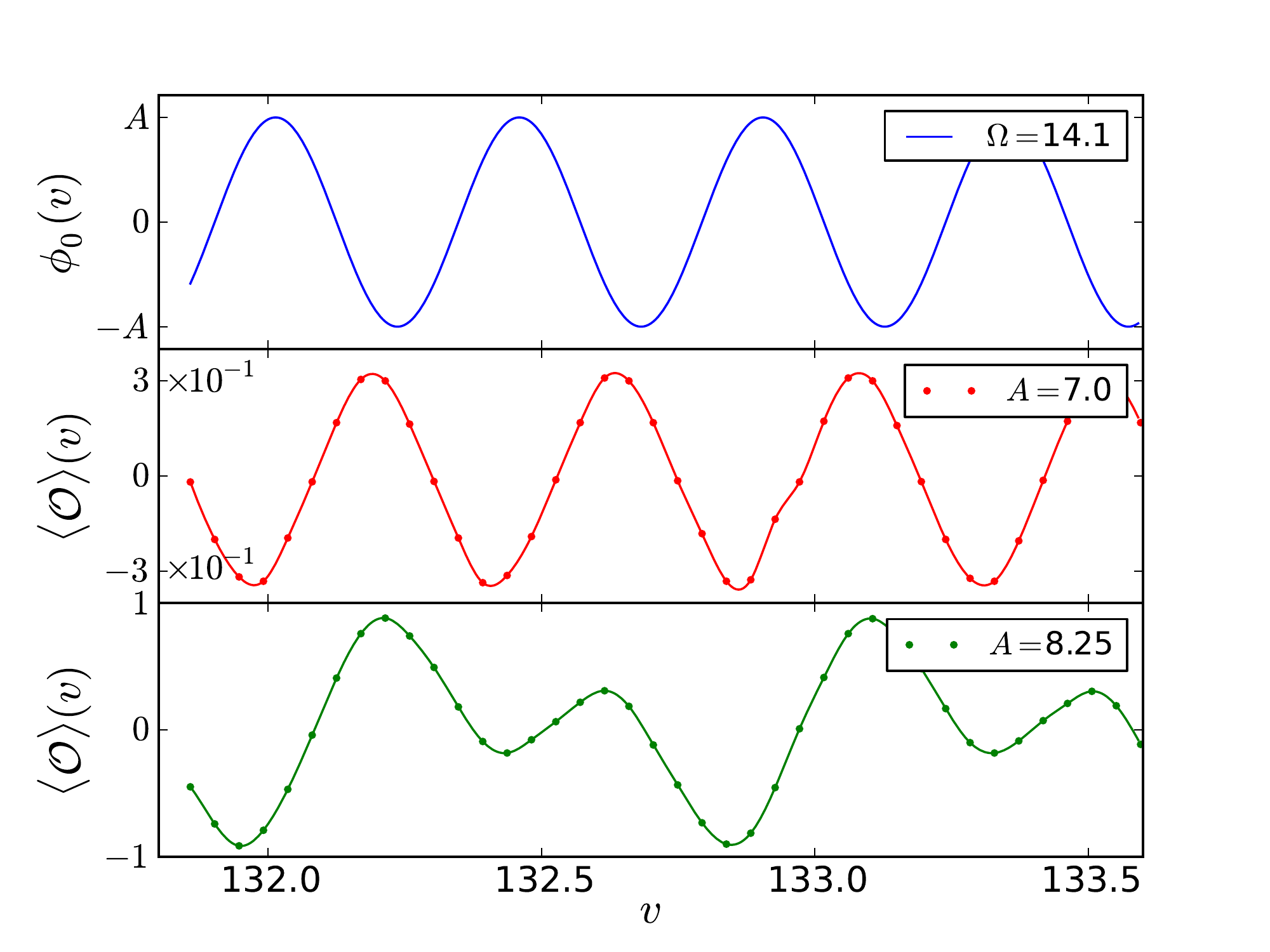}
\label{fig:timeplot}
}
\caption{Sinusoidal driving. In Fig.\,\ref{fig:profile} we see the late-time profile of the bulk field sampled after every period. Two distinct profiles indicate that the field takes two different configurations which repeat after every two periods. In Fig.\,\ref{fig:timeplot} we look at the time dependence of $\langle\mco\rangle$. The driving (upper blue) is at $\Omega=14.1$. For a smaller amplitude $A=7.0$ (middle red), the response is at the driving frequency (1-cycle). For a larger amplitude $A=8.25$ (lower green), the response is at half the driving frequency (2-cycle).}
\label{fig:proftime}
\end{figure}

\begin{figure}[h!]
\centering
\subfigure[Phase portrait showing a $2$-cycle.]{
\includegraphics[scale=0.38]{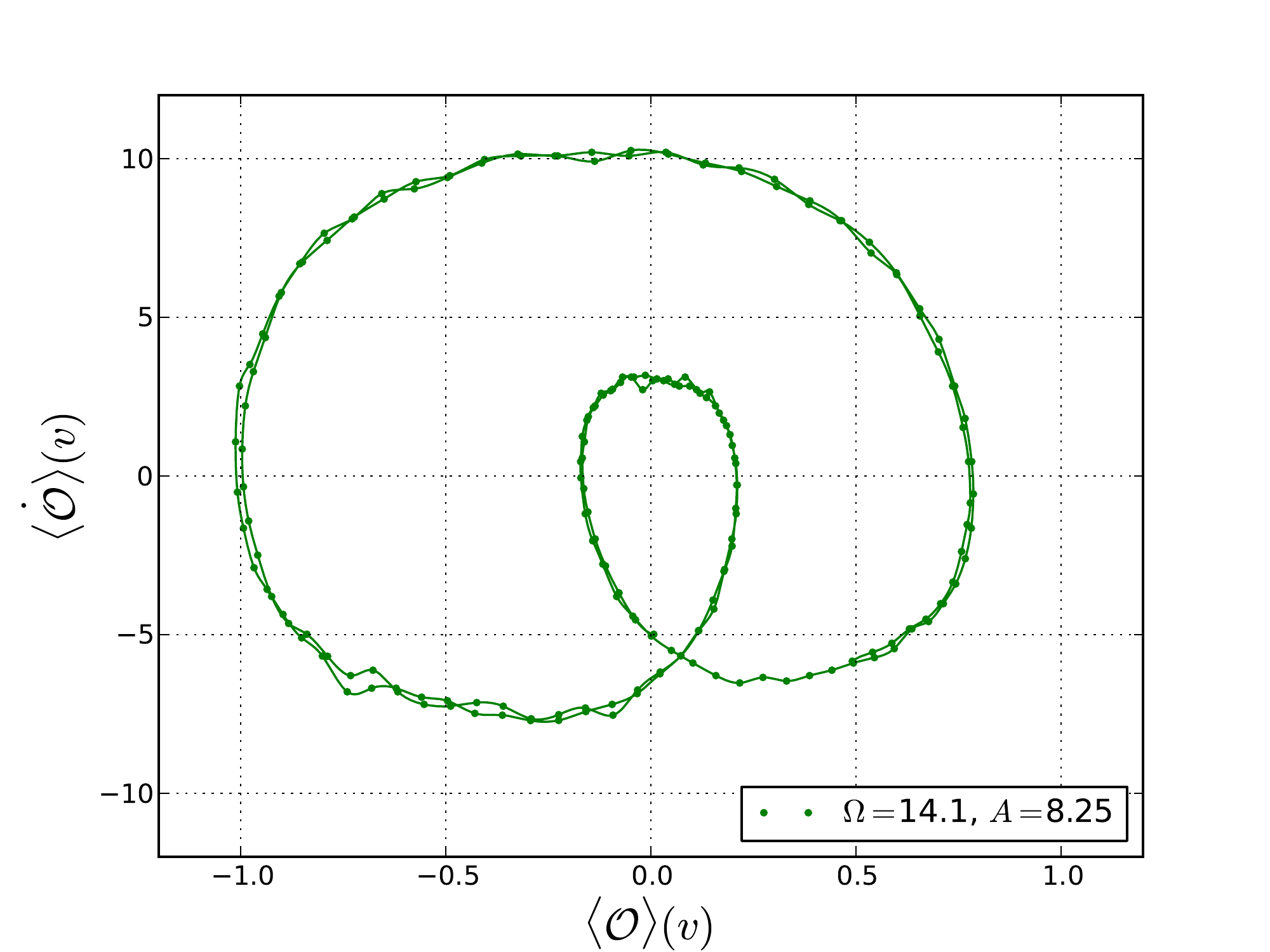}
\label{fig:poincare}
}
\subfigure[Bifurcation diagram.]{
\includegraphics[scale=0.38]{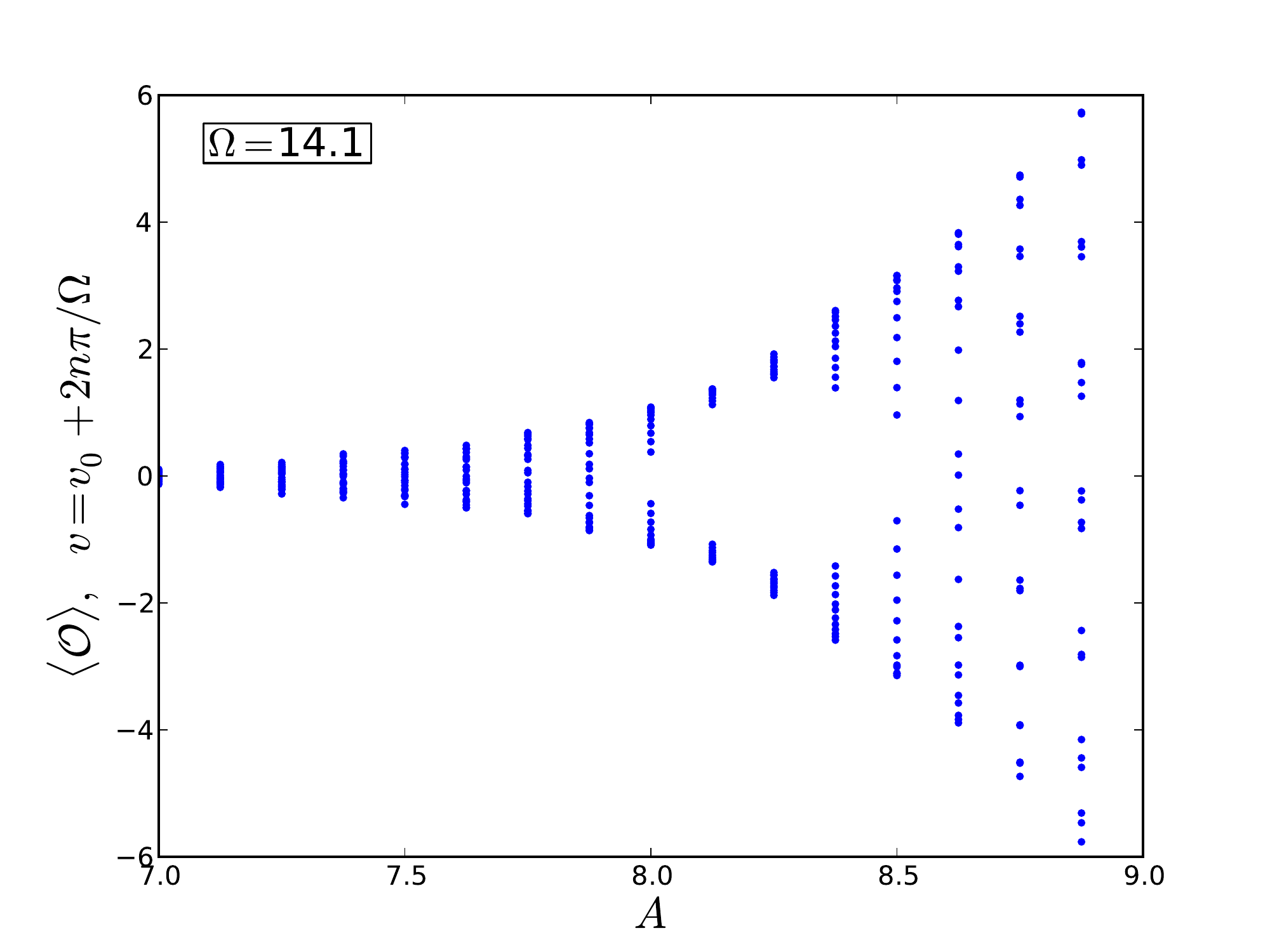}
\label{fig:bifdiag}
}
\caption{Transition to chaos. In Fig.\,\ref{fig:poincare}, the phase portrait has two lobes indicating a 2-cycle. In Fig.\,\ref{fig:bifdiag}, we see a period-doubling bifurcation and transition to chaos.}
\label{fig:poinbif}
\end{figure}

As we turn on nonlinearity parameter the response $\langle \hat{\mco} \rangle$ ceases to be sinusoidal, linear function of amplitude. However the response is still periodic.  We get novel nonlinear steady-state solutions. Depending on the initial conditions, there is  a transient behavior, which at late times usually settles into a steady limit cycle. The results are shown in Figures \ref{fig:proftime} and \ref{fig:poinbif}. For $\Omega=14.1$ and $A=7.0$, for example, the response is at the driving frequency -- the limit cycle is a 1-cycle. Typical profiles of the bulk field when the boundary value crosses zero (for the late time response) are shown in Fig.\,\ref{fig:profile}. The time-dependent response of the operator $\langle\mco\rangle=-\frac{\partial\phi}{\partial\rho}$ is plotted for two parameter values in Fig.\,\ref{fig:timeplot}.

\begin{figure}[h!]
\begin{center}
\includegraphics[scale=0.38]{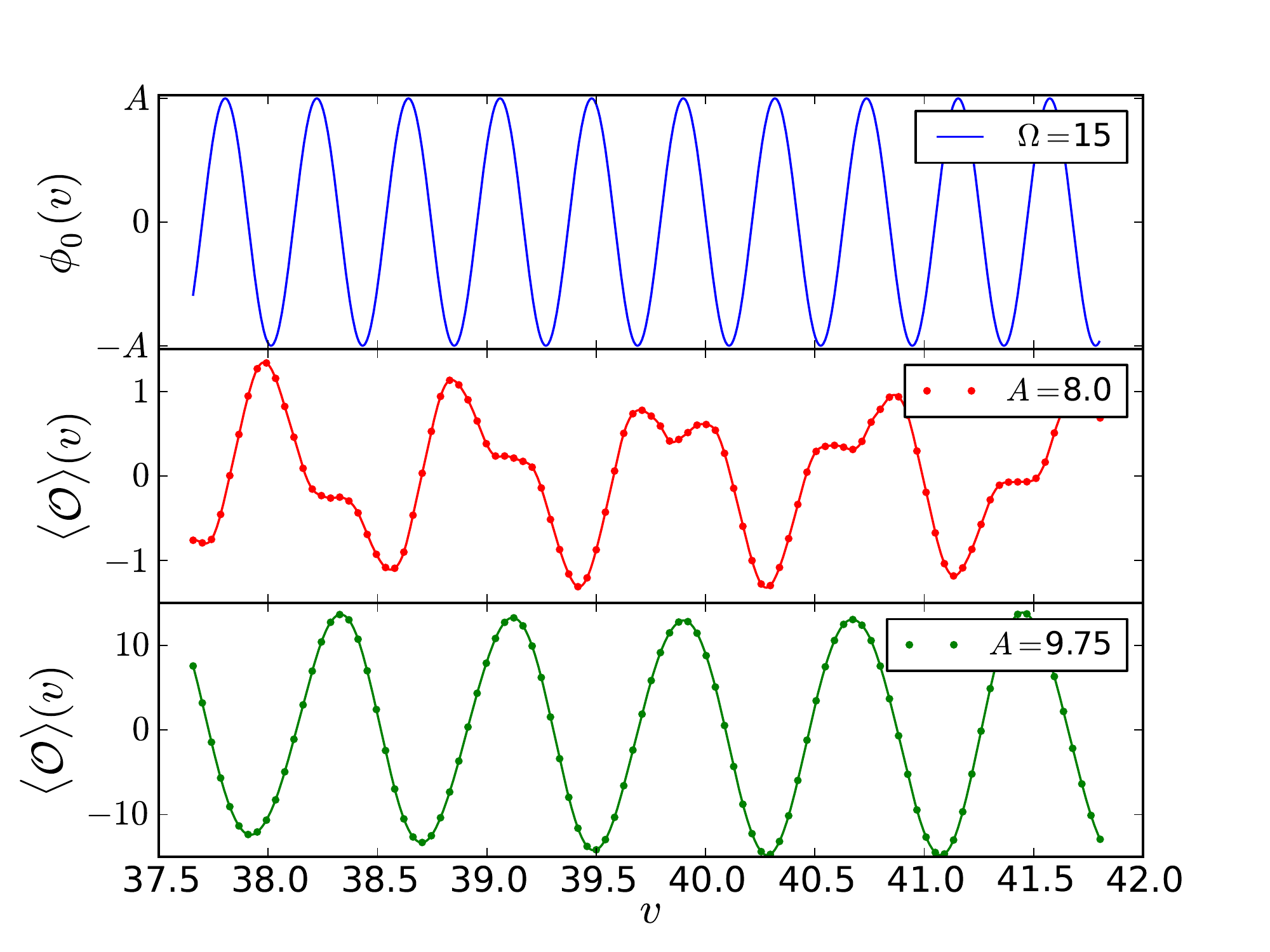}
\end{center}
\caption{Transition from chaos back to approximate order as the value of the amplitude is increased. For $A=8$, we see a chaotic behavior (middle red). For $A=9.75$, we see a roughly periodic behavior again.}
\label{fig:pastchaos}
\end{figure}

As the amplitude $A$ is increased keeping the frequency $\Omega$ fixed, we see a change in the qualitative behavior of the late-time response. For the same frequency $\Omega=14.1$, if the amplitude is increased to $A=8.25$, the alternate crests and troughs have different heights -- we see a 2-cycle. Indications of this are also seen in Fig.\,\ref{fig:profile}, where we see two different field profiles for alternate periods. The phase portrait for these parameters [Fig.\,\ref{fig:poincare}] has two distinct lobes clearly showing a 2-cycle. The system has thus undergone a period-doubling bifurcation between $A=7.0$ and $A=8.25$. From the bifurcation diagram [Fig.\,\ref{fig:bifdiag}], as the amplitude is increased keeping $\Omega=14.1$, we see a clear bifurcation near $A\approx8.0$. For larger values of amplitude $A\approx9$ we see a transition to complete chaos. As we increase the amplitude $A$ further, we end up again in a state which is ``less'' chaotic and looks quasiperiodic [Fig.\,\ref{fig:pastchaos}]. We also 
find approximate odd periods. This not unlike what we see in a simple damped pendulum. Our system is a $1\!+\!1$ dimensional system with an infinite number of degrees of freedom; it is unlikely that our chaos to order transition would be sharp and clean. Also one can question whether we ever get to an exact strange attractor configuration. It seems that the entire configuration space is explored once the system becomes chaotic. Further resolution of these questions needs more intensive numerics. 

It should be noted that the value of $A$ at which chaotic motion sets in varies inversely with $\Omega$.  At a small value of $\Omega \ll 1$, the system shows an adiabatic behavior even for large $A$. At small $\Omega$, the reverse is true and chaotic motion sets in for smaller $A$. 

\subsection{Conclusion and Quenching Dynamics}
In this work we have discussed the effect of a periodic disturbance of an $AdS$ black hole by scalar field sources. Using holography, this has been mapped to a disturbance of strongly coupled plasma by periodic sources. Our system has both dissipation (at the black hole horizon) and nonlinearity. It is interesting that we have found the standard features in nonlinear dynamics like the limit point and the periodic doubling route to chaos. Questions regarding the existence of strange attractors remain open. In this process, we also have numerically constructed novel solutions with holographic scalar fields. 

\begin{figure}[h!]
\centering
\subfigure[Slow driving.]{
\includegraphics[scale=0.38]{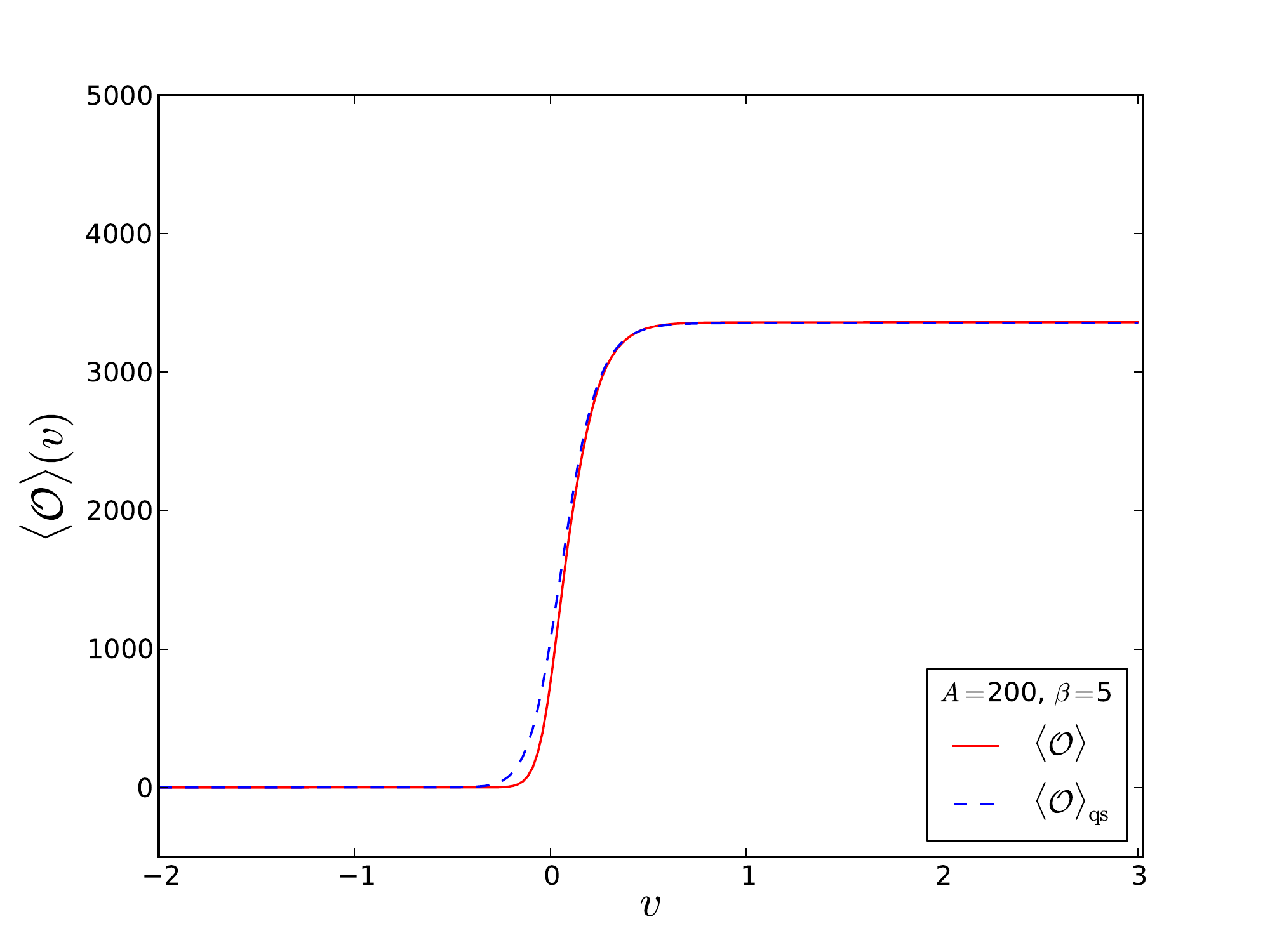}
\label{fig:slow}
}
\subfigure[Fast driving.]{
\includegraphics[scale=0.38]{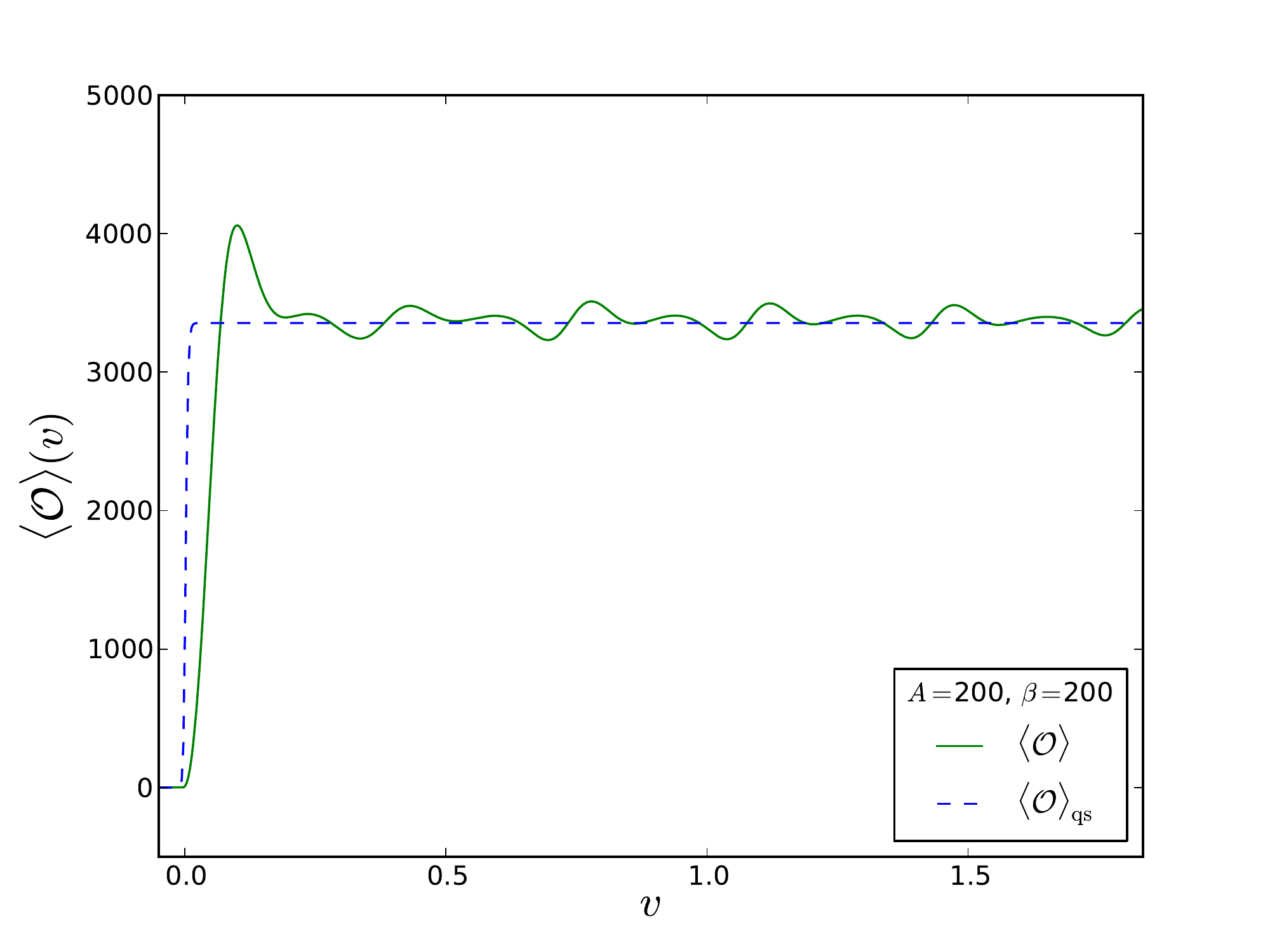}
\label{fig:fast}
}
\caption{Response of $\langle\mco\rangle$ for tanh driving. For a slow driving [Fig.\,\ref{fig:slow}], $\langle\mco\rangle$ essentially follows its quasistatic value (dashed blue). For a fast driving [Fig.\,\ref{fig:fast}], there are oscillations about the quasistatic value.}
\label{fig:quench}
\end{figure}

One may also choose many different kinds of driving other than what we have considered in this work. One important class of dynamics is quenching dynamics, where the time dependence of the disturbance dies off in the far past and far future. One can choose a ``tanh'' driving, i.e. $\phi_0(v)=A\tanh\beta v$ (see, for example, \cite{Basu:2012gg}) or a Gaussian pulse \cite{Bhaseen:2012gg}, both of which have been used in the context of holography. Our results for ``tanh'' driving with nonlinearity are shown in Figure~\ref{fig:quench}. For a very slow driving $\beta\ll\!1$, the response of the operator $\langle\mco\rangle$ is quasistatic -- the expectation value at any time is same as the expectation value for the same $\phi_0$ for the static case. For a slow variation ($\beta=\!5$), there is a slight lag but $\langle\mco\rangle$ essentially follows its quasistatic value $\langle\mco\rangle_{\rm \! qs}$ [Fig.\,\ref{fig:slow}]. For a faster variation ($\beta=\!200$), we see an oscillatory behavior about the quasi-
static value in the response [Fig.\,\ref{fig:fast}]. Our results have some of the features of the results of \cite{Basu:2012gg}; however there are crucial differences due to the presence of dissipation in our system\footnote{In \cite{Basu:2012gg}, the background is an $AdS$ soliton, which is nondissipative due to the absence of a horizon. Figure~2 of \cite{Basu:2012gg} has wiggles that do not die off and Fig.5 has oscillations that are sustained at a steady amplitude.}.

\section*{Acknowledgements}
This work is partially supported by National Science Foundation grants PHY-0970069, PHY-0855614 and PHY-1214341. We would like to thank Leopoldo Pando Zayas and Diptarka Das for discussion and collaboration during the initial stages of the work and Al Shapere and Sumit Das for valuable discussion. A.G. would like to thank Sayantani Bhattacharyya, Samriddhi Sankar Ray and especially Shailesh Lal for a many valuable discussions in the later stages of the work nearing its completion.

\appendix

\phantomsection
\addcontentsline{toc}{section}{References}
\bibliographystyle{JHEP}
\bibliography{inspirerefs,otherrefs}
\end{document}